\documentclass[12pt, preprint]{aastex}
\usepackage{natbib}
\usepackage{graphicx}
\usepackage{pifont}

\def\ion#1#2{#1 \textsc{#2}}
\begin{document}
\title{Fe IX Calculations for the Solar Dynamics Observatory}
\author{Adam R. Foster and Paola Testa}
\affil{Smithsonian Astrophysical Observatory, 60 Garden Street, Cambridge, MA 02138}
\email{afoster@cfa.harvard.edu}
\begin{abstract}
New calculations of the energy levels, radiative transition rates and collisional excitation rates of \ion{Fe}{ix} have been carried out using the Flexible Atomic Code, paying close attention to experimentally identified levels and extending existing calculations to higher energy levels. For lower levels, R-matrix collisional excitation rates from earlier work have been used. Significant emission is predicted by these calculations in the 5f-3d transitions, which will impact analysis of SDO AIA observations using the 94\AA\ filter.
\end{abstract}
\section{Introduction}
The launch of the Solar Dynamic Observatory (SDO) allows observation of the Sun in unprecedented detail. The Atmospheric Imaging Assembly (AIA, \citet{lemen2011}) provides multiple simultaneous images of the solar disk every 12 seconds, taken through a variety of narrowband filters centered on individual emission lines of interest. One such filter in centered on 94\AA, and targets both the \ion{Fe}{xviii} (93.923\AA) transition from $2\mbox{s}^1\ 2\mbox{p}^6\ ^2\mbox{S}_{1/2}\rightarrow 2\mbox{s}^2\ 2\mbox{p}^5\ ^2\mbox{P}_{3/2}$ at high temperatures and the \ion{Fe}{x} (94.012\AA) line from the $3\mbox{p}^4\ 4\mbox{s}^1\ ^2\mbox{D}_{5/2}\rightarrow 3\mbox{p}^5\ ^2\mbox{P}_{3/2}$ transition in cooler plasma. The emission from these two lines occurs in very different temperature ranges and therefore can be distinguished if the plasma temperature distribution is known \citep{boerner2011}.


\citet{lepson2002} observed spectra of \ion{Fe}{vii} to \ion{Fe}{x} using an electron beam ion trap (EBIT) and a grazing incidence spectrometer with resolution of $\approx300$ at 100\AA\ to observe lines in the 60 to 140\AA\ range. They estimated that 70\% of the emission in this band was unaccounted for by the existing atomic data in the \textsc{mekal} \citep{kaastramewe1993, mewe1995} database. Since then, none of the major atomic databases (e.g. \textsc{chianti} \citep{dere2009}, AtomDB \citep{foster2011}, \textsc{nist} \citep{ralchenko2011}, and the successor to \textsc{mekal}, \textsc{spex} \citep{kaastra1996}) have updated their atomic data for \ion{Fe}{ix} to include any lines in this region. Of particular interest are two \ion{Fe}{ix} lines, observed at 93.59 and 94.07\AA, which both fall in the bandpass of the 94\AA\ AIA filter.

These lines were identified by comparing the EBIT results with calculations using the Hebrew University Lawrence Livermore Atomic Code (\textsc{hullac}, \citet{barshalom2001}) as belonging to the $3\mbox{p}^5 5\mbox{f}^1 \rightarrow 3\mbox{p}^5 3\mbox{d}^1$ transitions of \ion{Fe}{ix}. The structure and collisional calculations of the current version of \textsc{chianti}, version 6.0.1 (and, by extension, AtomDB which uses the \textsc{chianti} data for this ion) only include configurations up to the n=4 principal quantum shell, and therefore omit these transitions. The NIST \citep{ralchenko2011} database does include some n=5 transitions but not these lines or levels.

In this work, we have carried out calculations using the Flexible Atomic Code (FAC, \citet{gu2003}) to extend the energy level and collisional calculations to include higher energy levels up to the n=6 shell, merged the resulting data with the best available collisional and radiative data for lower levels, and used the collisional-radiative code \textsc{apec} \citep{smith2001} to model the resulting emission and therefore the effect on the AIA 94\AA\ filter flux.

\section{Method}

The NIST atomic spectra database \citep{ralchenko2011} lists 35 observed energy levels for \ion{Fe}{ix}, including the lowest 17 energy levels and 18 others up to the 3p$^5$ 5s$^1$ level. Solar observations using the HINODE/EIS instrument have led to the identification of four more energy levels: $3\mbox{p}^4 3\mbox{d}^2\ {}^3\mbox{G}_{\left[4,5,3\right]}$ and $3\mbox{p}^5 4\mbox{p}^1\ {}^1\mbox{S}_0$  \citep{young2009}.

\begin{deluxetable}{lccclccc}
\tablecolumns{8} 
\tablewidth{0pc} 
\tablecaption{The configurations included in previous caclulations of energy levels, A-values and collision strengths of \ion{Fe}{ix}. S2002 denotes the \citet{storey2002} data while A2006 denotes \citet{aggarwal2006}. Levels marked ``CI'' are included only for configuration interaction purposes.\label{tab:configs} } 
\tablehead{ Config. & S2002 & A2006 & Current & Config. & S2002 & A2006 & Current}
\startdata
3s$^2$ 3p$^6$               &  Y  &  Y & Y  &
  3s$^2$ 3p$^3$ 3d$^3$        &  CI &  Y & Y  \\
3s$^2$ 3p$^5$ 3d$^1$        &  Y  &  Y & Y  &
  3s$^2$ 3p$^2$ 3d$^4$        &     &    & CI \\
3s$^2$ 3p$^5$ 4s$^1$        &  Y  &  Y & Y  &
  3s$^1$ 3p$^6$ 3d$^1$        &  Y  &  Y & Y  \\
3s$^2$ 3p$^5$ 4p$^1$        &  Y  &  Y & Y  &
  3s$^1$ 3p$^6$ 4l$^1$        &     &  Y & Y  \\
3s$^2$ 3p$^5$ 4d$^1$        &  CI &  Y & Y  &
  3s$^1$ 3p$^6$ 5l$^1$        &     &    & CI \\
3s$^2$ 3p$^5$ 4f$^1$        &     &  Y & Y  &
  3s$^1$ 3p$^6$ 6l$^1$        &     &    & CI \\
3s$^2$ 3p$^5$ 5l$^1$        &     &    & Y  &
  3s$^1$ 3p$^5$ 3d$^2$        &  CI &  Y & Y  \\
3s$^2$ 3p$^5$ 6l$^1$        &     &    & Y  &
  3s$^1$ 3p$^5$ 3d$^1$ 4l$^1$ &     &  Y & CI \\
3s$^2$ 3p$^4$ 3d$^2$        &  Y  &  Y & Y  &
  3s$^1$ 3p$^5$ 3d$^1$ 5l$^1$ &     &    & CI \\
3s$^2$ 3p$^4$ 3d$^1$ 4l$^1$ &     &  Y & Y  &
  3s$^1$ 3p$^5$ 3d$^1$ 6l$^1$ &     &    & CI \\
3s$^2$ 3p$^4$ 3d$^1$ 5l$^1$ &     &    & CI &
  3s$^1$ 3p$^4$ 3d$^3$        &  CI &  Y & CI \\
3s$^2$ 3p$^4$ 3d$^1$ 6l$^1$ &     &    & CI &
  3p$^6$ 3d$^2$               &  CI &  Y & CI \\
3s$^2$ 3p$^4$ 4p$^2$        &     &    & CI &
  3p$^5$ 3d$^3$               &  CI &    & CI \\
\enddata
\end{deluxetable}

Theoretical calculations of the \ion{Fe}{ix} structure have been performed by several groups. The configurations included in some of these are listed in Table \ref{tab:configs}. \citet{storey2002} performed \textsc{superstructure} \citep{eissner1974} calculations of the first 140 energy levels combined with an R-Matrix collision calculation. \citet{aggarwal2006} revisited this using the General Purpose Relativistic Atomic Structure Package (GRASP, \citet{dyall1989}) and FAC to calculate energy levels and transition rates. They used various configuration combinations in order to match the first 17 energy levels as closely as possible, while also paying particular attention to the first 360 energy levels. They found that the effect of the CI between many of the configurations on the energy levels and subsequent oscillator strength calculations is of importance, in particular the $3\mbox{s}^2 3\mbox{p}^3 3\mbox{d}^3$. This was omitted during another set of calculations by \citet{verma2006} using the \textsc{civ3} \citep{hibbert1975} code, and gave very different results compared to those of \citet{aggarwal2006} and observed energy levels. The configurations listed in Table \ref{tab:configs} are those from their best fitting GRASP run.

\begin{figure}
\begin{center}
\includegraphics[width=10cm]{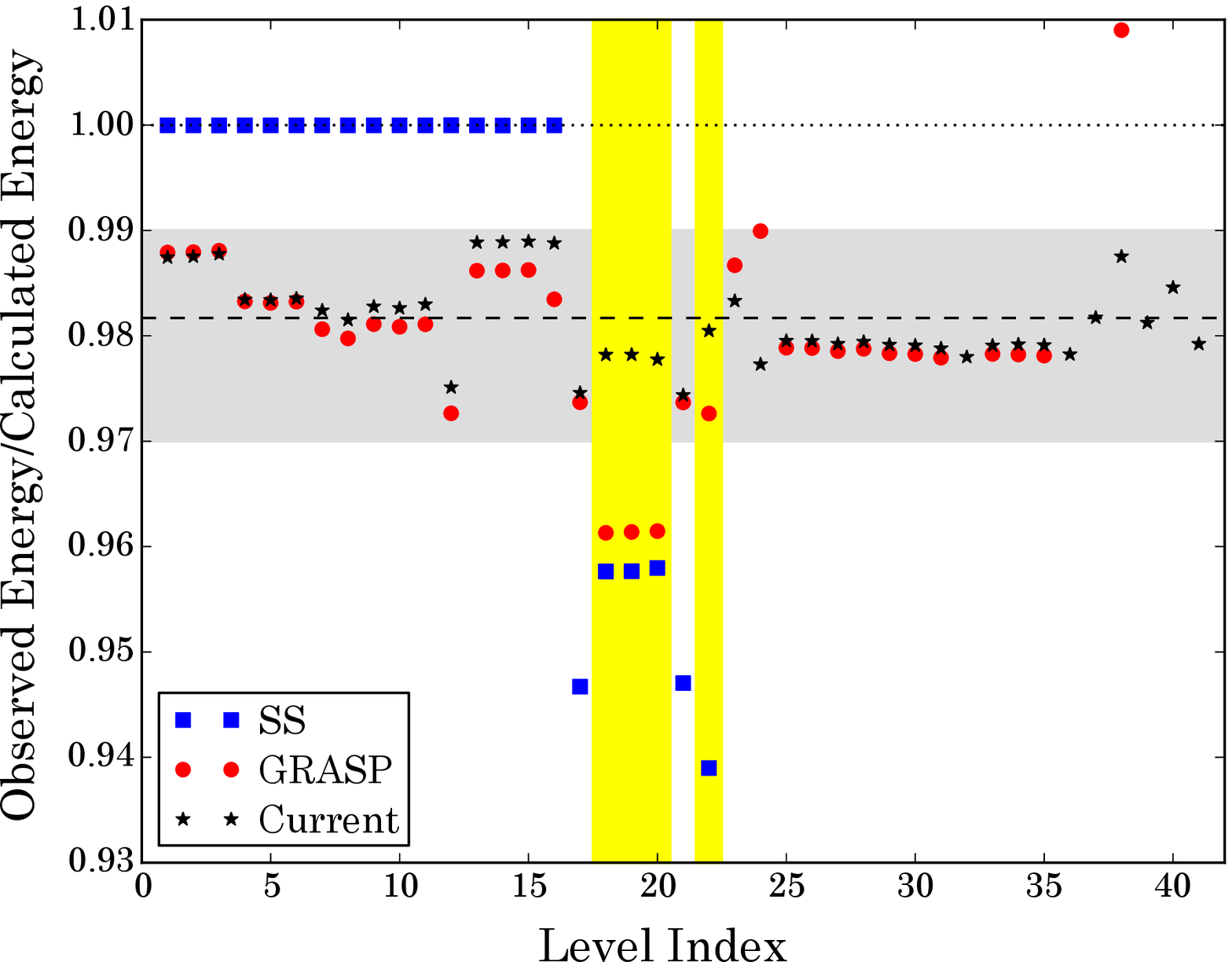}
\caption{\label{fig:energy_corr}The ratio of the observed and calculated energies for \ion{Fe}{ix} levels from a variety of methods. Squares: \textsc{superstructure} \citep{storey2002}; circles: GRASP \citep{aggarwal2006}; stars: this work. The levels with observed energy levels from \textsc{hinode} observations \citep{young2009} are highlighted, as is the 1\% to 3\% difference band in which all the current results fall. The dashed line indicates the correction factor used for all higher energy levels. A description of these levels is in Table \ref{tab:energies}, with their indexes in column 5. Only those levels with observed counterparts are included.}
\end{center}
\end{figure}

The energies of \citet{storey2002} agree with the observed energies to within 6\% (the energies of the first 17 levels were set to match observed values after the structure calculation was complete), while those of \citet{aggarwal2006} agree within the observed energies 1 to 3\%. The exception to this are the four levels from \citep{young2009}, numbers 94, 95, 96 and 144 in our energy ordering. These were not identified until after the \citet{aggarwal2006} data was created, and there is a substantial difference in both the energies and the level ordering caused by their introduction (see Figure \ref{fig:energy_corr}).

In our FAC calculation, we have experimented with different configuration sets to obtain a good match with the observed energies in both absolute energy and energy level ordering, and to include the higher $n$ shells from which the emission takes place. We have included the $3\mbox{s}^2 3\mbox{p}^5 5\mbox{l}^1$ and $3\mbox{s}^2 3\mbox{p}^5 6\mbox{l}^1$ configurations to include the higher $n$ emission which we are trying to characterize. 

We have also included many other configurations for their CI effects. The coupling between configurations of the same parity can affect the energy ordering of many of the levels during the structure calculation. In this case, most of these levels do not contribute observed emission lines, and their energies are much higher than those of immediate spectroscopic interest for which there are observed lines ($> 200$eV), leading them to straddle the ionization energy of the ion (i.e. many of them are auto-ionizing). To keep our calculation to a reasonable size, we have included many configurations only during the structure calculation for CI purposes and omitted them from the collisional and radiative calculations: again, these are listed in Table \ref{tab:configs}.

Of these additional configurations, it is the inclusion of the $3\mbox{s}^2 3\mbox{p}^2 3\mbox{d}^4$ and $3\mbox{s}^2 3\mbox{p}^4 4\mbox{p}^2$ configurations in the structure calculation which adds the CI necessary to bring the energies of the \citet{young2009} levels back to within 2.5\% of the observed values, and also creates an energy ordering which matches observations for all identified levels. Using our much larger FAC calculation, we obtain values comparable to those of \citet{aggarwal2006} for the lower energy levels and again within 1 to 2.6\% of observed values at higher levels. This does lead to significant changes in the energy ordering of levels when compared to the calculations of \citet{aggarwal2006}. Our energy levels, combined with our best attempts to identify matching levels have been listed in Table \ref{tab:energies}.

\begin{deluxetable}{cllllclc}
\tablecolumns{8} 
\tablewidth{0pc} 
\tablecaption{The list of energy levels resulting from this work. $E_{FAC}$ refers to the original results from FAC calculations, $E_{corr}$ are the energies after correction as described in the text. For each level with an observed energy value, the index of this level in Figure 1 is given ($Ind_{fig 1}$): for these levels $E_{corr} = E_{observed}$. For comparison, the \textsc{grasp} results of \citet{aggarwal2006} are listed ($E_G$), along with the energy order from that work $(\mbox{Ind}_{\mbox{\tiny{G}}})$. A star denotes a level for which a different configuration is found between our work and \citet{aggarwal2006}. \label{tab:energies}} 
\tablehead{
Index & 
\textit{jj} Symbol & 
$\mbox{E}_{\mbox{\tiny{FAC}}}$ & 
$\mbox{E}_{\mbox{\tiny{corr}}}$ & 
$\mbox{Ind}_{\mbox{\tiny{Fig.1}}}$ &
$\mbox{E}_{\mbox{\tiny{G}}}$&
$\mbox{Ind}_{\mbox{\tiny{G}}}$\\
&\colhead{} & (eV) & (eV) & \colhead{}& (eV) &\colhead{}}
\startdata
  1 & $3\mbox{p}_{\frac{1}{2}}^2 3\mbox{p}_{\frac{3}{2}}^4 \left(0,0\right)_{0}$                                                                                                                              &   0.000 &   0.000&   0 &   0.000 &   1 \\ 
  2 & $3\mbox{p}_{\frac{1}{2}}^2 3\mbox{p}_{\frac{3}{2}}^3 3\mbox{d}_{\frac{3}{2}}^1 \left(0,\frac{3}{2},\frac{3}{2}\right)_{0}$                                                                              &  50.949 &  50.309&   1 &  50.925 &   2 \\ 
  3 & $3\mbox{p}_{\frac{1}{2}}^2 3\mbox{p}_{\frac{3}{2}}^3 3\mbox{d}_{\frac{3}{2}}^1 \left(0,\frac{3}{2},\frac{3}{2}\right)_{1}$                                                                              &  51.264 &  50.625&   2 &  51.242 &   3 \\ 
  4 & $3\mbox{p}_{\frac{1}{2}}^2 3\mbox{p}_{\frac{3}{2}}^3 3\mbox{d}_{\frac{5}{2}}^1 \left(0,\frac{3}{2},\frac{5}{2}\right)_{2}$                                                                              &  51.923 &  51.288&   3 &  51.907 &   4 \\ 

 94 & $3\mbox{p}_{\frac{1}{2}}^2 3\mbox{p}_{\frac{3}{2}}^3 4\mbox{s}_{\frac{1}{2}}^1 \left(0,\frac{3}{2},\frac{1}{2}\right)_{1}$                                                                              & 120.922 & 117.850&  17 & 121.033 &  94 \\ 
 95 & $3\mbox{p}_{\frac{1}{2}}^2 3\mbox{p}_{\frac{3}{2}}^2 3\mbox{d}_{\frac{3}{2}}^1 3\mbox{d}_{\frac{5}{2}}^1 \left(0,2,\frac{3}{2},\frac{5}{2}\right)_{4}$                                                  & 121.280 & 118.637&  18 & 123.412 &  97 \\ 
 96 & $3\mbox{p}_{\frac{1}{2}}^1 3\mbox{p}_{\frac{3}{2}}^3 3\mbox{d}_{\frac{3}{2}}^1 3\mbox{d}_{\frac{5}{2}}^1 \left(\frac{1}{2},\frac{3}{2},\frac{3}{2},\frac{5}{2}\right)_{5}$                              & 121.346 & 118.703&  19 & 123.470 &  98 \\ 
 97 & $3\mbox{p}_{\frac{1}{2}}^2 3\mbox{p}_{\frac{3}{2}}^2 3\mbox{d}_{\frac{3}{2}}^1 3\mbox{d}_{\frac{5}{2}}^1 \left(0,2,\frac{3}{2},\frac{5}{2}\right)_{3}$                                                  & 121.461 & 118.761&  20 & 123.521 &  99 \\ 
 98 & $3\mbox{p}_{\frac{1}{2}}^1 3\mbox{p}_{\frac{3}{2}}^4 4\mbox{s}_{\frac{1}{2}}^1 \left(\frac{1}{2},0,\frac{1}{2}\right)_{0}$                                                                              & 122.173 & 119.247&     & 122.330 &  95 \\ 
 99 & $3\mbox{p}_{\frac{1}{2}}^1 3\mbox{p}_{\frac{3}{2}}^3 3\mbox{d}_{\frac{3}{2}}^1 3\mbox{d}_{\frac{5}{2}}^1 \left(\frac{1}{2},\frac{3}{2},\frac{3}{2},\frac{5}{2}\right)_{3}$                              & 122.577 & 119.521&     & 124.507 & 100 \\ 
100 & $3\mbox{p}_{\frac{1}{2}}^1 3\mbox{p}_{\frac{3}{2}}^4 4\mbox{s}_{\frac{1}{2}}^1 \left(\frac{1}{2},0,\frac{1}{2}\right)_{1}$                                                                              & 122.863 & 119.715&  21 & 122.951 &  96 \\ 
\enddata
\tablecomments{Table 2 is published in its entirety in the electronic edition of ApJL. A portion is shown here for guidance regarding its form and content.}
\end{deluxetable}

We note (see Figure \ref{fig:energy_corr}) that our calculated energies are without exception higher than the observed values. This offset scales simply with energy, resulting in calculated levels which are 1-2.5\% larger than observed values. Since the goal of these calculations is to produce useful spectra for astrophysical analysis, we have adjusted the level energies after the structure calculation to match the observed values. For intermediate levels, the multiplier has been interpolated in energy and applied. The $3\mbox{p}^5 5\mbox{f}^1$ levels of primary interest in this work are of a higher energy than any other experimentally identified levels of \ion{Fe}{ix}. Noting the relatively uniform overestimate of the energy in the structure calculations, we have scaled the energies for all levels above the highest observed energy level by the mean of the adjustment used for experimentally identifed levels, 0.9817. The observed energy levels are the 17 levels from NIST, the 4 levels from \citet{young2009}, and four levels calculated from the \citet{lepson2002} measurements: 
(3p$_{\frac{1}{2}}^2$ 3p$_{\frac{3}{2}}^3$ 4f$_{\frac{5}{2}}^1)_{2}$ (E=163.2eV), 
(3p$_{\frac{1}{2}}^1$ 3p$_{\frac{3}{2}}^4$ 4f$_{\frac{5}{2}}^1)_{2}$ (E=164.9eV), 
(3p$_{\frac{1}{2}}^2$ 3p$_{\frac{3}{2}}^3$ 5f$_{\frac{7}{2}}^1)_{2}$ (E=188.9eV), 
(3p$_{\frac{1}{2}}^1$ 3p$_{\frac{3}{2}}^4$ 5f$_{\frac{5}{2}}^1)_{2}$ (E=189.6eV).

Radiative rates were calculated by \citet{aggarwal2006} based on their structure calculation. We have initially attempted to use these radiative rates, however the significant disagreements in the energy level ordering have made this problematic for most levels. We have therefore used the \citet{aggarwal2006} values for transitions among the lowest 17 energy levels, where the ordering is definite, and the remainder have been calculated using the relativistic method within FAC.

\citet{storey2002} performed R-Matrix calculations, producing collisional excitation rates among the lowest 140 energy levels. Further work was performed using FAC by \citet{liang2009}, although again these did not incorporate n=5 configurations.

A full R-matrix calculation including all of the excited levels on the n=5 level would be a prohibitively large calculation and is beyond the scope of this work. We have therefore use the distorted wave FAC collision code to calculate collision strengths between all of the levels of \ion{Fe}{ix} included in our structure calculations. The advantage of this approach is the fast production of results and the inclusion of large numbers of levels. The downside is the omission of low-energy resonance effects, which can be significant. We therefore use the R-matrix collision strengths of \citet{storey2002} where they exist. We note that the problem of matching levels between the calculations persists. For most levels, we have matched the levels with the same electron configuration and total angular momentum, $J$, and then paired these results in energy order. For the $3\mbox{p}^4 3\mbox{d}^2$ and $3\mbox{p}^5 4\mbox{p}^1$ levels there are different numbers of levels (for the former configuration, our calculation has 2 more levels for each of $J$=2, 3 and 4; for the latter \citet{storey2002} have 2 more). We have therefore combined these two configurations, which overlap completely in energy at the higher end of the $3\mbox{p}^5 3\mbox{d}^2$ energies ($\approx 130$eV).

\section{Results}
\begin{figure}
\begin{center}
\includegraphics[width=16cm]{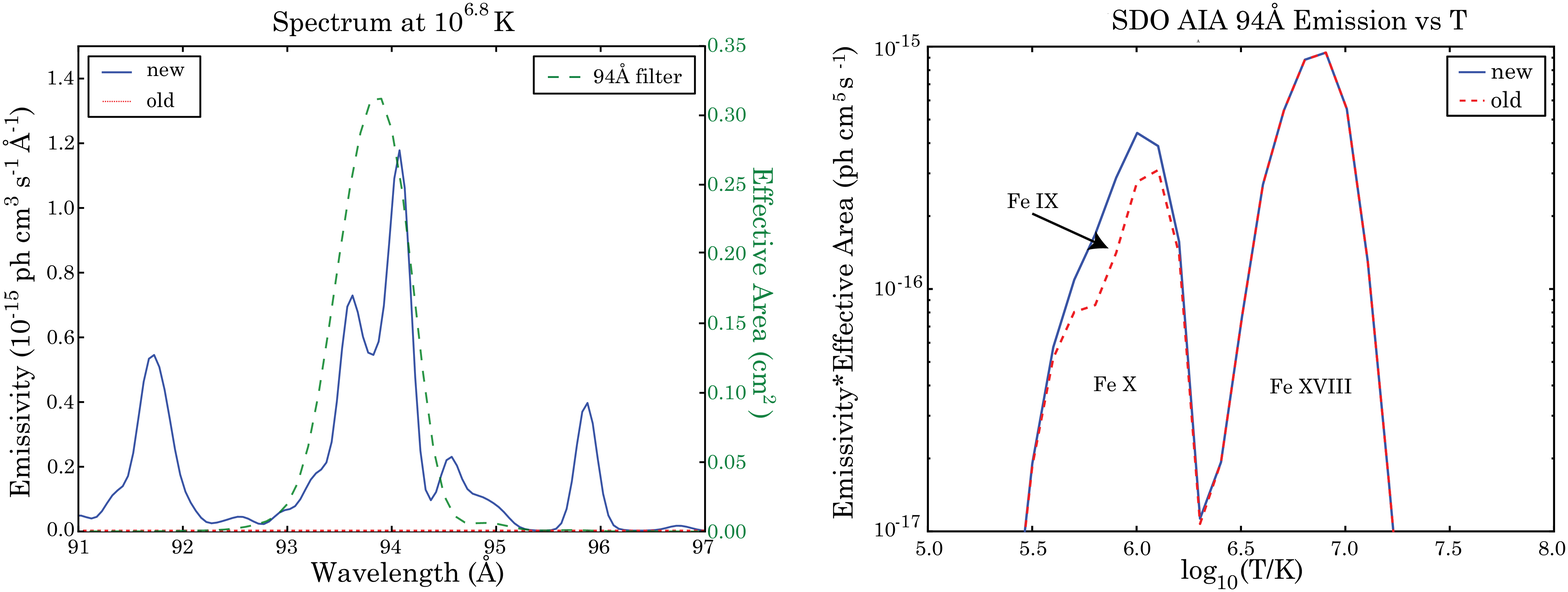}
\caption{\label{fig:fe9_spec}\textit{Left}: the emissivity of \ion{Fe}{ix} at $T=10^6$K. Dotted line: from AtomDB v2.0.1; solid line: from AtomDB v2.0.1 with \ion{Fe}{ix} data from this work; dashed line: the effective area of the 94\AA\ channel of the SDO/AIA. The AtomDB v2.0.1 data is effectively zero, and hence too small to be notable on the graph. \textit{Right}: The total emissivitiy of elemental emission lines convolved with the effective area of the 94\AA\ filter. Dashed line: data from AtomDB v2.0; solid line: data from AtomDB v2.0 including the \ion{Fe}{ix} data from this work.}
\end{center}
\end{figure}

We have taken the results of our FAC calculation, merged where appropriate with other calculations as outlined above, and used the \textsc{apec} code to model the emission from an optically thin plasma with solar photospheric elemental abundances \citep{andersgrevesse1989} in collisional ionization equilibrium. We show this result in Figure \ref{fig:fe9_spec}, for T=$10^6$K plasma with the emission lines broadened by Gaussians with width 0.05\AA. Overplotted on this figure is the same result from the currently version (2.0) of AtomDB \citep{foster2011}, which uses the old \ion{Fe}{ix} data. Overlaid is the effective area curve of the AIA instrument with the 94\AA\ filter in place, taken from \citet{boerner2011}, which can also be obtained by the routine aia\_get\_response.pro in the SolarSoftware package. It can be seen that there are clearly two strong lines and several weaker lines, which are also from 5f$\rightarrow$3d transitions, within the region of interest. 

The weaker lines are a significant fraction of the two main 5f-3d lines, with emissivities of around 10-20\% of the main lines. Their exact wavelengths are, however, unknown: the 5f-3d lines were already weak in the EBIT measurements, the weaker neighboring lines were not distinguished from the background. The correction applied to the 5f-3d lines was a further $\pm 5\%$ compared to the simple multiplication by the 0.9817 scaling factor, in opposite directions. This implies that a correction at a similar level may be required in the case of these weaker lines. We have not further adjusted the energy levels of these weaker lines after the general initial scaling. Further experimental measurements of these lines would be valuable in estimating their effects more clearly.

In Figure \ref{fig:fe9_spec} we have also convolved the line emissivities for all ions with the effective area of the filter at each wavelength, and show the total as a function of temperature. We have done this for the AtomDB 2.0.1 model, which omits the 94\AA\ lines, and for an identical dataset but using the new \ion{Fe}{ix} data from this work. The low temperature peak, previously due to \ion{Fe}{x}, is significantly increased due to the strong \ion{Fe}{ix} emission as part of this work, by over a factor of 2 at $\log_{10}(T_e/K)=5.8$.

\begin{figure}
\begin{center}
\includegraphics[width=10cm]{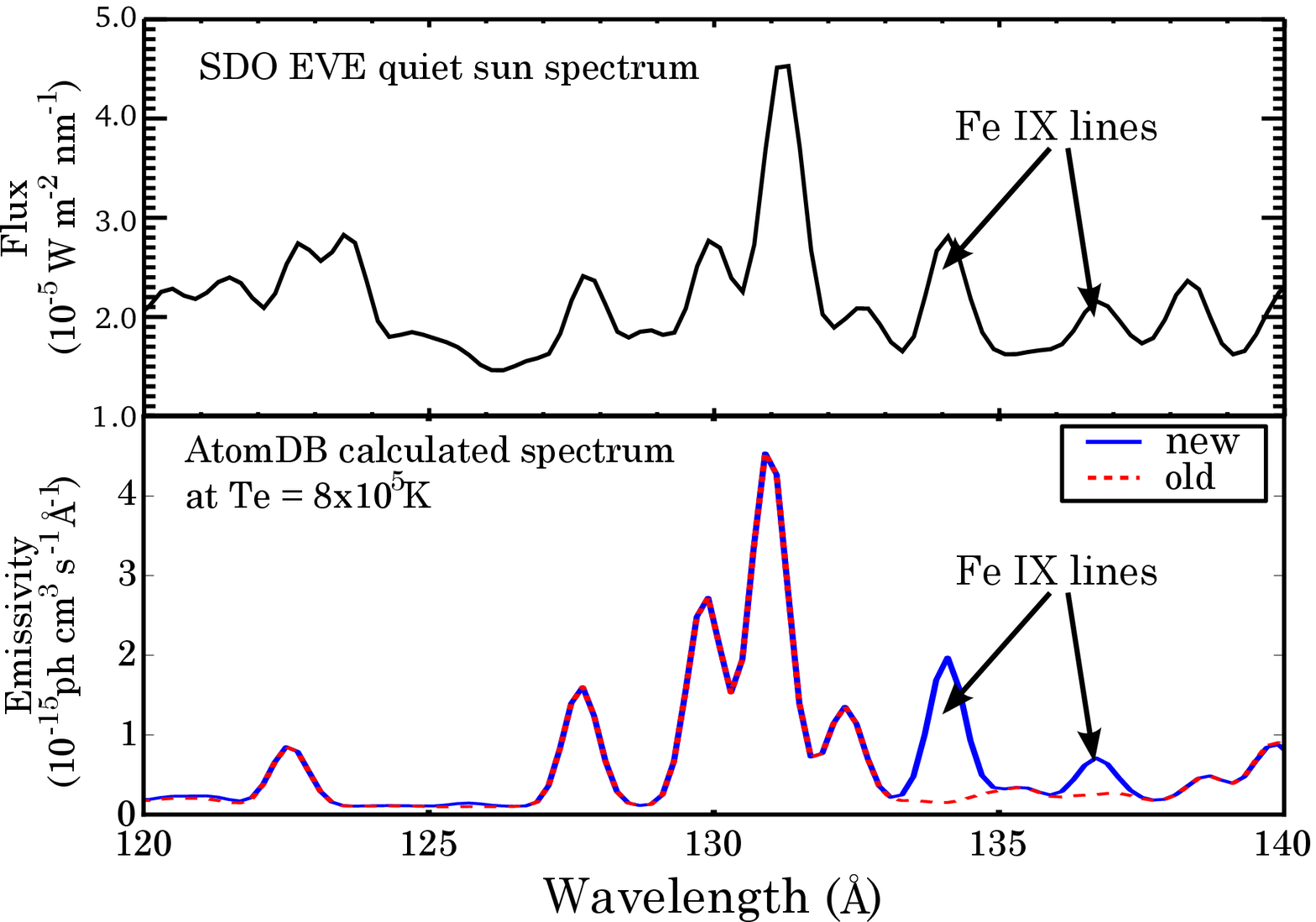}
\caption{\label{fig:131a}\textit{Top}: A quiet Sun spectrum taken from the SDO EVE MEGS instrument. \textit{Bottom}: The spectrum calculated using AtomDB 2.0.1, with (dashed) the old \ion{Fe}{ix} data and (solid) the new data from this work, with all lines broadened by Gaussians with $\sigma=0.2$\AA.}
\end{center}
\end{figure}

In addition, further lines are positively identified in the \ion{Fe}{ix} spectrum. The lines at 134.08 and 136.70\AA, identified as belonging to the $3\mbox{p}^5 4\mbox{f}^1 \rightarrow 3\mbox{p}^5 3\mbox{d}^1$ transition, are clearly identifiable in the new calculations of the spectrum. Figure ~\ref{fig:131a} shows the spectrum calculated using the old and new \ion{Fe}{ix} data, combined with a quiet Sun spectrum taken from the SDO Extreme Ultraviolet Variability Experiment MEGS. The previously unidentified lines at 134.08\AA\ and 136.70\AA\ are clearly observable. These wavelengths fall sufficiently far away from the 131\AA\ filter transmission band that there is no discernable change in the estimated flux in this filter due to the inclusion of these lines.

\section{Discussion}
\label{sec:discussion}
The lack of information on the \ion{Fe}{ix} emission in the 94\AA\ region has been well known since the work of \citet{lepson2002}. \citet{testa2011} compared the existing atomic data in \textsc{chianti} 6.0.1 with observations of Procyon using the Chandra LETG spectrometer, which covers wavelengths up to $\sim 170$\AA. They noted in these observations a significant missing flux in the 94\AA\ region. Since the launch of the SDO, the lack of a collisional excitation calculation has led to several authors using creative means to account for this difference. When looking at coronal loop emission \citet{schmelz2011} used a scaling factor obtained by comparing the relative intensities of the known transition lines and the $5\mbox{f} \rightarrow 3\mbox{d}$ in the spectrum observed by \citet{lepson2002}. They eventually discounted this band from their analysis due to a low count rate, so the effectiveness of this method is unknown. \citet{aschwanden2011}, when looking at the temperature structure of coronal loops, derived a ``correction factor'', $q_{94}$, for the low temperature ($\log_{10}(T/K) \le 6.3$) response function of the 94\AA\ filter, $R_{94}$, by fitting their 100 results in the other filters with $q_{94}$ a free parameter. They obtained $q_{94} = 6.7 \pm 1.7$ at temperatures around $\log_{10}(T/K)=6.0$. The results from this work imply a much smaller correction factor, with $q_{94}\approx 2$ being the maximum value at $\log_{10}(T/K) = 5.8$, and $\approx 1.25$ at $\log_{10}(T/K) = 6.0$. There is significant discrepancy here: it would be an interesting exercise to investigate whether the DEM analysis of \citet{aschwanden2011} can be self consistent using the new emission estimate in this filter band. It is also possible that there are futher lines from different ions in the filter which have not yet been correctly handled.

It is difficult to estimate uncertainties in the results from this work due to the many overlapping sources of error. As well as the two main lines which are the focus of this work, there are  many smaller lines of uncertain wavlength. Given the observed and initially calculated 5f $\rightarrow$ 3d wavelengths differ by  around 0.5\%, adding normally distributed random errors to these lines of $\pm 0.5$\AA\ gives a 10\% standard deviation in the resulting \ion{Fe}{ix} emission in the 94\AA\ filter. This 10\% fluctuation is then approximately a 5\% effect on the total emission in the band once \ion{Fe}{x} is also included. Given that the estimates of uncertainty in distorted wave excitation collisions are usually not better than 20\%, this is not expected to be a dominating source of error in the calculation.

For those levels which appear in both this work and the collisional calculation of \citet{storey2002}, comparison of the collision strengths for excitation from the ground state at $T=10^6$K between this data and our FAC results show that they vary by on average around 25\%. This is a very approximate lower limit on the likely errors on the calculation of emission in the 94\AA\ band, since excluding cascades from the n=6 shell, this is the only way to populated the 5f upper levels in this model, and for these levels we are using the distorted wave, not R-matrix, methods.

This data has been incorporated into AtomDB, to be released fully in AtomDB v2.1.0 which is due for release later in 2011. In the meantime, the data can be obtained from the AtomDB website, \url{www.atomdb.org}.

\acknowledgments
ARF acknowledges funding from NASA ADP grant \#NNX09AC71G. PT has been supported by contract SP02H1701R from Lockheed-Martin to SAO.


\begin{thebibliography}{99}
\bibitem[Aggarwal et al.(2006)]{aggarwal2006}Aggarwal, K. M., Keenan, F. P., Kato, T., and Murakamia, I., 2006, A\&A, 460, 331
\bibitem[Anders \& Grevesse(1989)]{andersgrevesse1989}Anders, E., and Grevesse, N., 1989, Geochimica et Cosmochimica Acta, 53, 197
\bibitem[Aschwanden \& Boerner(2011)]{aschwanden2011}Aschwanden, M. J., and Boerner, P., 2011, ApJ, 732, 81
\bibitem[Bar Shalom, Klapish \& Oreg(2001)]{barshalom2001}Bar-Shalom, A., Klapisch, M., and Oreg, J. 2001, JQSRT, 71, 169
\bibitem[Boerner et al.(2011)]{boerner2011}Boerner, P., et al., 2011, to appear in Solar Physics, http://www.lmsal.com/sdodocs/doc?cmd=dcur\&proj\_num=SDOD0063\&file\_type=pdf
\bibitem[Dere et al.(2009)]{dere2009}Dere, K. P. et al. 2009, A\&A, 498, 915
\bibitem[Dyall(1989)]{dyall1989}Dyall, K. G., Grant, I. P, Johnson, C. T., Parpia, F. A., and Plummer, E. P., 1989, CPC, 55, 424
\bibitem[Eissner, Jones \& Nussbaumer(1974)]{eissner1974}Eissner, W., Jones, M., and Nussbaumer, H., 1974, CPC 8, 270
\bibitem[Foster et al.(2011)]{foster2011}Foster, A. R., et al., in prep.
\bibitem[Gu(2003)]{gu2003}Gu, M. F., 2003, ApJ, 582, 1241
\bibitem[Hibbert(1975)]{hibbert1975}Hibbert, A., 1975, CPC, 9, 141
\bibitem[Kaastra \& Mewe(1993)]{kaastramewe1993}Kaastra, J. S., and Mewe, R., 1993, Legacy, 3, 6
\bibitem[Kaastra, Mewe \& Nieuwenhuijzen(1996)]{kaastra1996}Kaastra, J. S., Mewe, R., and Nieuwenhuijzen, H., 1996, UV and X-ray Spectroscopy of Astrophysical and Laboratory Plasmas, 411
\bibitem[Lemen et al.(2011)]{lemen2011}Lemen, J., et al., 2011, Solar Physics, 10.1007/s11207-011-9776-8.
\bibitem[Lepson et al.(2002)]{lepson2002}Lepson, J. K., et al., 2002, ApJ, 578, 648
\bibitem[Liang et al.(2009)]{liang2009}Liang, G. Y., et al., 2009, ApJ, 702, 838
\bibitem[Mewe, Kaastra \& Liedahl(1995)]{mewe1995}Mewe, R., Kaastra, J. S., and Liedahl, D. A., 1995, Legacy, 6, 16 
\bibitem[Testa et al (2011)]{testa2011}Testa, P., et al, 2011, submitted to ApJ.
\bibitem[Ralchenko et al.(2011)]{ralchenko2011} Ralchenko, Yu., Kramida, A.E., Reader, J., and NIST ASD Team (2011). NIST Atomic Spectra Database (ver. 4.1.0), [Online]. Available: http://physics.nist.gov/asd3  [2011, June 15]. National Institute of Standards and Technology, Gaithersburg, MD. 
\bibitem[Schmelz et al.(2011)]{schmelz2011}Schmelz, J. T., et al., 2011, ApJ, 731, 49
\bibitem[Smith et al.(2001)]{smith2001}Smith, R. K., Brickhouse, N. S., Liedahl, D. A., and Raymond, J. C., 2001, ApJL, 556, 91
\bibitem[Storey, Zeippen \& Le Dourneuf(2002)]{storey2002}Storey, P. J., Zeippen, C. J., and Le Dourneuf, M., 2002, A\&A, 394, 753
\bibitem[Verma et al.(2006)]{verma2006}Verma, N., Jha, A. K. S., and Mohan, M., 2006, ApJS, 164, 297
\bibitem[Young(2009)]{young2009}Young, P. R., 2009, ApJL, 691, 77
\end{thebibliography}
\end{document}